\documentstyle{eqwidths}

\def\PsfigVersion{1.9}
\ifx\undefined\psfig\else\endinput\fi

\let\LaTeXAtSign=\@
\let\@=\relax
\edef\psfigRestoreAt{\catcode`\@=\number\catcode`@\relax}
\catcode`\@=11\relax
\newwrite\@unused
\def\ps@typeout#1{{\let\protect\string\immediate\write\@unused{#1}}}
\ps@typeout{psfig/tex \PsfigVersion}

\def\figurepath{./}

\def\@nnil{\@nil}
\def\@empty{}
\def\@psdonoop#1\@@#2#3{}
\def\@psdo#1:=#2\do#3{\edef\@psdotmp{#2}\ifx\@psdotmp\@empty \else
    \expandafter\@psdoloop#2,\@nil,\@nil\@@#1{#3}\fi}
\def\@psdoloop#1,#2,#3\@@#4#5{\def#4{#1}\ifx #4\@nnil \else
       #5\def#4{#2}\ifx #4\@nnil \else#5\@ipsdoloop #3\@@#4{#5}\fi\fi}
\def\@ipsdoloop#1,#2\@@#3#4{\def#3{#1}\ifx #3\@nnil 
       \let\@nextwhile=\@psdonoop \else
      #4\relax\let\@nextwhile=\@ipsdoloop\fi\@nextwhile#2\@@#3{#4}}
\def\@tpsdo#1:=#2\do#3{\xdef\@psdotmp{#2}\ifx\@psdotmp\@empty \else
    \@tpsdoloop#2\@nil\@nil\@@#1{#3}\fi}
\def\@tpsdoloop#1#2\@@#3#4{\def#3{#1}\ifx #3\@nnil 
       \let\@nextwhile=\@psdonoop \else
      #4\relax\let\@nextwhile=\@tpsdoloop\fi\@nextwhile#2\@@#3{#4}}
\ifx\undefined\fbox
\newdimen\fboxrule
\newdimen\fboxsep
\newdimen\ps@tempdima
\newbox\ps@tempboxa
\long\def\fbox#1{\leavevmode\setbox\ps@tempboxa\hbox{#1}\ps@tempdima\fboxrule
    \advance\ps@tempdima \fboxsep \advance\ps@tempdima \dp\ps@tempboxa
   \hbox{\lower \ps@tempdima\hbox
  {\vbox{\hrule height \fboxrule
          \hbox{\vrule width \fboxrule \hskip\fboxsep
          \vbox{\vskip\fboxsep \box\ps@tempboxa\vskip\fboxsep}\hskip 
                 \fboxsep\vrule width \fboxrule}
                 \hrule height \fboxrule}}}}
\fi
\newread\ps@stream
\newif\ifnot@eof       
\newif\if@noisy        
\newif\if@atend        
\newif\if@psfile       
{\catcode`\%=12\global\gdef\epsf@start{
\def\epsf@PS{PS}
\def\epsf@getbb#1{%
\openin\ps@stream=#1
\ifeof\ps@stream\ps@typeout{Error, File #1 not found}\else
   {\not@eoftrue \chardef\other=12
    \def\do##1{\catcode`##1=\other}\dospecials \catcode`\ =10
    \loop
       \if@psfile
	  \read\ps@stream to \epsf@fileline
       \else{
	  \obeyspaces
          \read\ps@stream to \epsf@tmp\global\let\epsf@fileline\epsf@tmp}
       \fi
       \ifeof\ps@stream\not@eoffalse\else
       \if@psfile\else
       \expandafter\epsf@test\epsf@fileline:. \\%
       \fi
          \expandafter\epsf@aux\epsf@fileline:. \\%
       \fi
   \ifnot@eof\repeat
   }\closein\ps@stream\fi}%
\long\def\epsf@test#1#2#3:#4\\{\def\epsf@testit{#1#2}
			\ifx\epsf@testit\epsf@start\else
\ps@typeout{Warning! File does not start with `\epsf@start'.  It may not be a PostScript file.}
			\fi
			\@psfiletrue} 
{\catcode`\%=12\global\let\epsf@percent=
\long\def\epsf@aux#1#2:#3\\{\ifx#1\epsf@percent
   \def\epsf@testit{#2}\ifx\epsf@testit\epsf@bblit
	\@atendfalse
        \epsf@atend #3 . \\%
	\if@atend	
	   \if@verbose{
		\ps@typeout{psfig: found `(atend)'; continuing search}
	   }\fi
        \else
        \epsf@grab #3 . . . \\%
        \not@eoffalse
        \global\no@bbfalse
        \fi
   \fi\fi}%
\def\epsf@grab #1 #2 #3 #4 #5\\{%
   \global\def\epsf@llx{#1}\ifx\epsf@llx\empty
      \epsf@grab #2 #3 #4 #5 .\\\else
   \global\def\epsf@lly{#2}%
   \global\def\epsf@urx{#3}\global\def\epsf@ury{#4}\fi}%
\def\epsf@atendlit{(atend)} 
\def\epsf@atend #1 #2 #3\\{%
   \def\epsf@tmp{#1}\ifx\epsf@tmp\empty
      \epsf@atend #2 #3 .\\\else
   \ifx\epsf@tmp\epsf@atendlit\@atendtrue\fi\fi}

\chardef\psletter = 11 
\chardef\other = 12

\newif \ifdebug 
\newif\ifc@mpute 
\c@mputetrue 

\let\then = \relax
\def\r@dian{pt }
\let\r@dians = \r@dian
\let\dimensionless@nit = \r@dian
\let\dimensionless@nits = \dimensionless@nit
\def\internal@nit{sp }
\let\internal@nits = \internal@nit
\newif\ifstillc@nverging
\def \Mess@ge #1{\ifdebug \then \message {#1} \fi}

{ 
	\catcode `\@ = \psletter
	\gdef \nodimen {\expandafter \n@dimen \the \dimen}
	\gdef \term #1 #2 #3%
	       {\edef \t@ {\the #1}
		\edef \t@@ {\expandafter \n@dimen \the #2\r@dian}%
		\t@rm {\t@} {\t@@} {#3}%
	       }
	\gdef \t@rm #1 #2 #3%
	       {{%
		\count 0 = 0
		\dimen 0 = 1 \dimensionless@nit
		\dimen 2 = #2\relax
		\Mess@ge {Calculating term #1 of \nodimen 2}%
		\loop
		\ifnum	\count 0 < #1
		\then	\advance \count 0 by 1
			\Mess@ge {Iteration \the \count 0 \space}%
			\Multiply \dimen 0 by {\dimen 2}%
			\Mess@ge {After multiplication, term = \nodimen 0}%
			\Divide \dimen 0 by {\count 0}%
			\Mess@ge {After division, term = \nodimen 0}%
		\repeat
		\Mess@ge {Final value for term #1 of 
				\nodimen 2 \space is \nodimen 0}%
		\xdef \Term {#3 = \nodimen 0 \r@dians}%
		\aftergroup \Term
	       }}
	\catcode `\p = \other
	\catcode `\t = \other
	\gdef \n@dimen #1pt{#1} 
}

\def \Divide #1by #2{\divide #1 by #2} 

\def \Multiply #1by #2
       {{
	\count 0 = #1\relax
	\count 2 = #2\relax
	\count 4 = 65536
	\Mess@ge {Before scaling, count 0 = \the \count 0 \space and
			count 2 = \the \count 2}%
	\ifnum	\count 0 > 32767 
	\then	\divide \count 0 by 4
		\divide \count 4 by 4
	\else	\ifnum	\count 0 < -32767
		\then	\divide \count 0 by 4
			\divide \count 4 by 4
		\else
		\fi
	\fi
	\ifnum	\count 2 > 32767 
	\then	\divide \count 2 by 4
		\divide \count 4 by 4
	\else	\ifnum	\count 2 < -32767
		\then	\divide \count 2 by 4
			\divide \count 4 by 4
		\else
		\fi
	\fi
	\multiply \count 0 by \count 2
	\divide \count 0 by \count 4
	\xdef \product {#1 = \the \count 0 \internal@nits}%
	\aftergroup \product
       }}

\def\r@duce{\ifdim\dimen0 > 90\r@dian \then   
		\multiply\dimen0 by -1
		\advance\dimen0 by 180\r@dian
		\r@duce
	    \else \ifdim\dimen0 < -90\r@dian \then  
		\advance\dimen0 by 360\r@dian
		\r@duce
		\fi
	    \fi}

\def\Sine#1%
       {{%
	\dimen 0 = #1 \r@dian
	\r@duce
	\ifdim\dimen0 = -90\r@dian \then
	   \dimen4 = -1\r@dian
	   \c@mputefalse
	\fi
	\ifdim\dimen0 = 90\r@dian \then
	   \dimen4 = 1\r@dian
	   \c@mputefalse
	\fi
	\ifdim\dimen0 = 0\r@dian \then
	   \dimen4 = 0\r@dian
	   \c@mputefalse
	\fi
	\ifc@mpute \then
		\divide\dimen0 by 180
		\dimen0=3.141592654\dimen0
		\dimen 2 = 3.1415926535897963\r@dian 
		\divide\dimen 2 by 2 
		\Mess@ge {Sin: calculating Sin of \nodimen 0}%
		\count 0 = 1 
		\dimen 2 = 1 \r@dian 
		\dimen 4 = 0 \r@dian 
		\loop
			\ifnum	\dimen 2 = 0 
			\then	\stillc@nvergingfalse 
			\else	\stillc@nvergingtrue
			\fi
			\ifstillc@nverging 
			\then	\term {\count 0} {\dimen 0} {\dimen 2}%
				\advance \count 0 by 2
				\count 2 = \count 0
				\divide \count 2 by 2
				\ifodd	\count 2 
				\then	\advance \dimen 4 by \dimen 2
				\else	\advance \dimen 4 by -\dimen 2
				\fi
		\repeat
	\fi		
			\xdef \sine {\nodimen 4}%
       }}

\def\Cosine#1{\ifx\sine\UnDefined\edef\Savesine{\relax}\else
		             \edef\Savesine{\sine}\fi
	{\dimen0=#1\r@dian\advance\dimen0 by 90\r@dian
	 \Sine{\nodimen 0}
	 \xdef\cosine{\sine}
	 \xdef\sine{\Savesine}}}	      

\def\psdraft{
	\def\@psdraft{0}
}
\def\psfull{
	\def\@psdraft{100}
}

\psfull

\newif\if@scalefirst
\def\psscalefirst{\@scalefirsttrue}
\def\psrotatefirst{\@scalefirstfalse}
\psrotatefirst

\newif\if@draftbox
\def\psnodraftbox{
	\@draftboxfalse
}
\def\psdraftbox{
	\@draftboxtrue
}
\@draftboxtrue

\newif\if@prologfile
\newif\if@postlogfile
\def\pssilent{
	\@noisyfalse
}
\def\psnoisy{
	\@noisytrue
}
\psnoisy
\newif\if@bbllx
\newif\if@bblly
\newif\if@bburx
\newif\if@bbury
\newif\if@height
\newif\if@width
\newif\if@rheight
\newif\if@rwidth
\newif\if@angle
\newif\if@clip
\newif\if@verbose
\newif\if@scale
\def\@p@@sclip#1{\@cliptrue}

\newif\if@decmpr

\def\@p@@sfigure#1{\def\@p@sfile{null}\def\@p@sbbfile{null}
	        \openin1=#1.bb
		\ifeof1\closein1
	        	\openin1=\figurepath#1.bb
			\ifeof1\closein1
			        \openin1=#1
				\ifeof1\closein1%
				       \openin1=\figurepath#1
					\ifeof1
					   \ps@typeout{Error, File #1 not found}
						\if@bbllx\if@bblly
				   		\if@bburx\if@bbury
			      				\def\@p@sfile{#1}%
			      				\def\@p@sbbfile{#1}%
							\@decmprfalse
				  	   	\fi\fi\fi\fi
					\else\closein1
				    		\def\@p@sfile{\figurepath#1}%
				    		\def\@p@sbbfile{\figurepath#1}%
						\@decmprfalse
	                       		\fi%
			 	\else\closein1%
					\def\@p@sfile{#1}
					\def\@p@sbbfile{#1}
					\@decmprfalse
			 	\fi
			\else
				\def\@p@sfile{\figurepath#1}
				\def\@p@sbbfile{\figurepath#1.bb}
				\@decmprtrue
			\fi
		\else
			\def\@p@sfile{#1}
			\def\@p@sbbfile{#1.bb}
			\@decmprtrue
		\fi}

\def\@p@@sfile#1{\@p@@sfigure{#1}}

\def\@p@@sbbllx#1{
		\@bbllxtrue
		\dimen100=#1
		\edef\@p@sbbllx{\number\dimen100}
}
\def\@p@@sbblly#1{
		\@bbllytrue
		\dimen100=#1
		\edef\@p@sbblly{\number\dimen100}
}
\def\@p@@sbburx#1{
		\@bburxtrue
		\dimen100=#1
		\edef\@p@sbburx{\number\dimen100}
}
\def\@p@@sbbury#1{
		\@bburytrue
		\dimen100=#1
		\edef\@p@sbbury{\number\dimen100}
}
\def\@p@@sheight#1{
		\@heighttrue
		\dimen100=#1
   		\edef\@p@sheight{\number\dimen100}
}
\def\@p@@swidth#1{
		\@widthtrue
		\dimen100=#1
		\edef\@p@swidth{\number\dimen100}
}
\def\@p@@srheight#1{
		\@rheighttrue
		\dimen100=#1
		\edef\@p@srheight{\number\dimen100}
}
\def\@p@@srwidth#1{
		\@rwidthtrue
		\dimen100=#1
		\edef\@p@srwidth{\number\dimen100}
}
\def\@p@@sangle#1{
		\@angletrue
		\edef\@p@sangle{#1} 
}
\def\@p@@srotate#1{\@p@@sangle{-#1}}
\def\@p@@sscale#1{
		\@scaletrue
		\edef\@p@sscale{#1}
}
\def\@p@@ssilent#1{ 
		\@verbosefalse
}
\def\@p@@sprolog#1{\@prologfiletrue\def\@prologfileval{#1}}
\def\@p@@spostlog#1{\@postlogfiletrue\def\@postlogfileval{#1}}
\def\@cs@name#1{\csname #1\endcsname}
\def\@setparms#1=#2,{\@cs@name{@p@@s#1}{#2}}
\def\ps@init@parms{
		\@bbllxfalse \@bbllyfalse
		\@bburxfalse \@bburyfalse
		\@heightfalse \@widthfalse
		\@rheightfalse \@rwidthfalse
		\@scalefalse
		\def\@p@sbbllx{}\def\@p@sbblly{}
		\def\@p@sbburx{}\def\@p@sbbury{}
		\def\@p@sheight{}\def\@p@swidth{}
		\def\@p@srheight{}\def\@p@srwidth{}
		\def\@p@sangle{0}
		\def\@p@sfile{} \def\@p@sbbfile{}
		\def\@p@scost{10}
		\def\@sc{}
		\@prologfilefalse
		\@postlogfilefalse
		\@clipfalse
		\if@noisy
			\@verbosetrue
		\else
			\@verbosefalse
		\fi
}
\def\parse@ps@parms#1{
	 	\@psdo\@psfiga:=#1\do
		   {\expandafter\@setparms\@psfiga,}}
\newif\ifno@bb
\def\bb@missing{
	\if@verbose{
		\ps@typeout{psfig: searching \@p@sbbfile \space  for bounding box}
	}\fi
	\no@bbtrue
	\epsf@getbb{\@p@sbbfile}
        \ifno@bb \else \bb@cull\epsf@llx\epsf@lly\epsf@urx\epsf@ury\fi
}	
\def\bb@cull#1#2#3#4{
	\dimen100=#1 bp\edef\@p@sbbllx{\number\dimen100}
	\dimen100=#2 bp\edef\@p@sbblly{\number\dimen100}
	\dimen100=#3 bp\edef\@p@sbburx{\number\dimen100}
	\dimen100=#4 bp\edef\@p@sbbury{\number\dimen100}
	\no@bbfalse
}
\newdimen\p@intvaluex
\newdimen\p@intvaluey
\def\rotate@#1#2{{\dimen0=#1 sp\dimen1=#2 sp
		  \global\p@intvaluex=\cosine\dimen0
		  \dimen3=\sine\dimen1
		  \global\advance\p@intvaluex by -\dimen3
		  \global\p@intvaluey=\sine\dimen0
		  \dimen3=\cosine\dimen1
		  \global\advance\p@intvaluey by \dimen3
		  }}
\def\compute@bb{
		\no@bbfalse
		\if@bbllx \else \no@bbtrue \fi
		\if@bblly \else \no@bbtrue \fi
		\if@bburx \else \no@bbtrue \fi
		\if@bbury \else \no@bbtrue \fi
		\ifno@bb \bb@missing \fi
		\ifno@bb \ps@typeout{FATAL ERROR: no bb supplied or found}
			\no-bb-error
		\fi
		\count203=\@p@sbburx
		\count204=\@p@sbbury
		\advance\count203 by -\@p@sbbllx
		\advance\count204 by -\@p@sbblly
		\edef\ps@bbw{\number\count203}
		\edef\ps@bbh{\number\count204}
		\if@angle 
			\Sine{\@p@sangle}\Cosine{\@p@sangle}
	        	{\dimen100=\maxdimen\xdef\r@p@sbbllx{\number\dimen100}
					    \xdef\r@p@sbblly{\number\dimen100}
			                    \xdef\r@p@sbburx{-\number\dimen100}
					    \xdef\r@p@sbbury{-\number\dimen100}}
                        \def\minmaxtest{
			   \ifnum\number\p@intvaluex<\r@p@sbbllx
			      \xdef\r@p@sbbllx{\number\p@intvaluex}\fi
			   \ifnum\number\p@intvaluex>\r@p@sbburx
			      \xdef\r@p@sbburx{\number\p@intvaluex}\fi
			   \ifnum\number\p@intvaluey<\r@p@sbblly
			      \xdef\r@p@sbblly{\number\p@intvaluey}\fi
			   \ifnum\number\p@intvaluey>\r@p@sbbury
			      \xdef\r@p@sbbury{\number\p@intvaluey}\fi
			   }
			\rotate@{\@p@sbbllx}{\@p@sbblly}
			\minmaxtest
			\rotate@{\@p@sbbllx}{\@p@sbbury}
			\minmaxtest
			\rotate@{\@p@sbburx}{\@p@sbblly}
			\minmaxtest
			\rotate@{\@p@sbburx}{\@p@sbbury}
			\minmaxtest
			\edef\@p@sbbllx{\r@p@sbbllx}\edef\@p@sbblly{\r@p@sbblly}
			\edef\@p@sbburx{\r@p@sbburx}\edef\@p@sbbury{\r@p@sbbury}
		\fi
		\count203=\@p@sbburx
		\count204=\@p@sbbury
		\advance\count203 by -\@p@sbbllx
		\advance\count204 by -\@p@sbblly
		\edef\@bbw{\number\count203}
		\edef\@bbh{\number\count204}
}
\def\in@hundreds#1#2#3{\count240=#2 \count241=#3
		     \count100=\count240	
		     \divide\count100 by \count241
		     \count101=\count100
		     \multiply\count101 by \count241
		     \advance\count240 by -\count101
		     \multiply\count240 by 10
		     \count101=\count240	
		     \divide\count101 by \count241
		     \count102=\count101
		     \multiply\count102 by \count241
		     \advance\count240 by -\count102
		     \multiply\count240 by 10
		     \count102=\count240	
		     \divide\count102 by \count241
		     \count200=#1\count205=0
		     \count201=\count200
			\multiply\count201 by \count100
		 	\advance\count205 by \count201
		     \count201=\count200
			\divide\count201 by 10
			\multiply\count201 by \count101
			\advance\count205 by \count201
		     \count201=\count200
			\divide\count201 by 100
			\multiply\count201 by \count102
			\advance\count205 by \count201
		     \edef\@result{\number\count205}
}
\def\ps@scaleinhundreds#1{
		\in@hundreds{#1}{\@p@sscale}{100}
		\edef#1{\@result}
}
\def\compute@wfromh{
		\in@hundreds{\@p@sheight}{\@bbw}{\@bbh}
		\edef\@p@swidth{\@result}
}
\def\compute@hfromw{
	        \in@hundreds{\@p@swidth}{\@bbh}{\@bbw}
		\edef\@p@sheight{\@result}
}
\def\compute@handw{
		\if@height 
			\if@width
			\else
				\compute@wfromh
			\fi
		\else 
			\if@width
				\compute@hfromw
			\else
				\edef\@p@sheight{\@bbh}
				\edef\@p@swidth{\@bbw}
			\fi
		\fi
}
\def\compute@resv{
		\if@rheight \else \edef\@p@srheight{\@p@sheight} \fi
		\if@rwidth \else \edef\@p@srwidth{\@p@swidth} \fi
}
\def\compute@sizes{
	\compute@bb
	\if@scalefirst\if@angle
	\if@width
	   \in@hundreds{\@p@swidth}{\@bbw}{\ps@bbw}
	   \edef\@p@swidth{\@result}
	\fi
	\if@height
	   \in@hundreds{\@p@sheight}{\@bbh}{\ps@bbh}
	   \edef\@p@sheight{\@result}
	\fi
	\fi\fi
	\compute@handw
	\compute@resv
	\if@scale
	   \if@verbose
	      \ps@typeout{(scaling by \@p@sscale)}%
	   \fi
	   \ps@scaleinhundreds{\@p@swidth}%
	   \ps@scaleinhundreds{\@p@sheight}%
	   \ps@scaleinhundreds{\@p@srwidth}%
	   \ps@scaleinhundreds{\@p@srheight}%
	\fi
}

\def\psfig#1{\vbox {
	\ps@init@parms
	\parse@ps@parms{#1}
	\compute@sizes
	\ifnum\@p@scost<\@psdraft{
		\special{ps::[begin] 	\@p@swidth \space \@p@sheight \space
				\@p@sbbllx \space \@p@sbblly \space
				\@p@sbburx \space \@p@sbbury \space
				startTexFig \space }
		\if@angle
			\special {ps:: \@p@sangle \space rotate \space} 
		\fi
		\if@clip{
			\if@verbose{
				\ps@typeout{(clip)}
			}\fi
			\special{ps:: doclip \space }
		}\fi
		\if@prologfile
		    \special{ps: plotfile \@prologfileval \space } \fi
		\if@decmpr{
			\if@verbose{
				\ps@typeout{psfig: including \@p@sfile.Z \space }
			}\fi
			\special{ps: plotfile "`zcat \@p@sfile.Z" \space }
		}\else{
			\if@verbose{
				\ps@typeout{psfig: including \@p@sfile \space }
			}\fi
			\special{ps: plotfile \@p@sfile \space }
		}\fi
		\if@postlogfile
		    \special{ps: plotfile \@postlogfileval \space } \fi
		\special{ps::[end] endTexFig \space }
		\vbox to \@p@srheight true sp{
			\hbox to \@p@srwidth true sp{
				\hss
			}
		\vss
		}
	}\else{
			\vbox to \@p@srheight true sp{
			\vss
			\hbox to \@p@srwidth true sp{\hss}
			\vss
			}

	}\fi
}}
\psfigRestoreAt
\let\@=\LaTeXAtSign

\newcommand{\beq}{\begin{equation}}
\newcommand{\eeq}{\end{equation}}
\newcommand{\beqnn}{\begin{displaymath}}	
\newcommand{\eeqnn}{\end{displaymath}}		
\newcommand{\beqa}{\begin{eqnarray}}
\newcommand{\eeqa}{\end{eqnarray}}
\newcommand{\beqann}{\begin{eqnarray*}}
\newcommand{\eeqann}{\end{eqnarray*}}
\newcommand{\nn}{\nonumber}
\newcommand{\ben}{\begin{enumerate}}
\newcommand{\een}{\end{enumerate}}
\newcommand{\bit}{\begin{itemize}}
\newcommand{\eit}{\end{itemize}}
\newcommand{\bc}{\begin{center}}
\newcommand{\ec}{\end{center}}

\newcommand{\tabref}[1]{Table~\ref{#1}}
\newcommand{\tabsref}[2]{Tables~\ref{#1} and~\ref{#2}}
\newcommand{\figref}[1]{Figure~\ref{#1}}
\newcommand{\figrefbare}[1]{\ref{#1}}
\newcommand{\figrefs}[1]{Figures~\ref{#1}}
\newcommand{\figsref}[2]{Figures~\ref{#1} and~\ref{#2}}
\newcommand{\eqref}[1]{Equation~\ref{#1}}
\newcommand{\eqsref}[2]{Equations~\ref{#1} and~\ref{#2}}
\newcommand{\chapref}[1]{Chapter~\ref{#1}}
\newcommand{\secref}[1]{Section~\ref{#1}}
\newcommand{\secsref}[2]{Sections~\ref{#1} and~\ref{#2}}
\newcommand{\appref}[1]{Appendix~\ref{#1}}

\newcommand{\Tabref}[1]{\tabref{#1}}
\newcommand{\Tabsref}[2]{\tabsref{#1}{#2}}
\newcommand{\Figref}[1]{\figref{#1}}
\newcommand{\Figrefs}[1]{\figrefs{#1}}
\newcommand{\Figsref}[2]{\figsref{#1}{#2}}
\newcommand{\Eqref}[1]{\eqref{#1}}
\newcommand{\Eqsref}[2]{\eqsref{#1}{#2}}
\newcommand{\Chapref}[1]{\chapref{#1}}
\newcommand{\Secref}[1]{\secref{#1}}
\newcommand{\Appref}[1]{\appref{#1}}

\newcommand{\equals}[1]{\mbox{$\stackrel{\rm #1}{=}$}}

\newcommand{\coude}{coud\'{e}}
\newcommand{\ie}{i.e.,}
\newcommand{\eg}{e.g.,}
\newcommand{\AIPS}{\mbox{$\cal AIPS$}}

\newcommand{\Htwo}{\mbox{H\,{\sc ii}}}

\newcommand{\eBoo}{\mbox{$\eta$~Boo}}
\newcommand{\aCen}{\mbox{$\alpha$~Cen}}

\newcommand{\about}{\mbox{$\sim$\,}}	
\newcommand{\degree}{\mbox{$^\circ$}}

\def\Msol{\mbox{${M}_\odot$}}
\def\Lsol{\mbox{${L}_\odot$}}
\def\Rsol{\mbox{${R}_\odot$}}
\def\gsol{\mbox{${g}_\odot$}}

\def\deg{\hbox{$^\circ$}}
\def\solar{\mbox{$_{\odot}$}}
\def\sun{\hbox{$\odot$}}
\def\earth{\hbox{$\oplus$}}
\def\la{\mathrel{\hbox{\rlap{\hbox{\lower4pt\hbox{$\sim$}}}\hbox{$<$}}}}
\def\ga{\mathrel{\hbox{\rlap{\hbox{\lower4pt\hbox{$\sim$}}}\hbox{$>$}}}}
\def\sq{\hbox{\rlap{$\sqcap$}$\sqcup$}}
\def\arcmin{\hbox{$^\prime$}}
\def\arcsec{\hbox{$^{\prime\prime}$}}
\def\fd{\hbox{$.\!\!^{\rm d}$}}
\def\fh{\hbox{$.\!\!^{\rm h}$}}
\def\fm{\hbox{$.\!\!^{\rm m}$}}
\def\fs{\hbox{$.\!\!^{\rm s}$}}
\def\fdg{\hbox{$.\!\!^\circ$}}
\def\farcm{\hbox{$.\mkern-4mu^\prime$}}
\def\farcs{\hbox{$.\!\!^{\prime\prime}$}}
\def\fp{\hbox{$.\!\!^{\scriptscriptstyle\rm p}$}}
\def\micron{\hbox{$\mu$m}}

\makeatletter 
 \def\sub#1{\relax\ifmmode _{\fam\z@ #1}\else
         $_{\fam\z@ #1}$\fi}
 \def\super#1{\relax\ifmmode ^{\fam\z@ #1}\else
         $^{\fam\z@ #1}$\fi}
\makeatother 

\newcommand{\downup}[3]{#1\sub{\rm #2}\super{\rm #3}}
\newcommand{\down}[2]{#1\sub{#2}}
\newcommand{\up}[2]{#1\super{#2}}

\newcommand{\comment}[1]{\relax}

\long\def\COMMENT#1\ENDCOMMENT{}
\def\ENDCOMMENT{}

\newcommand{\Aosc}{\down{A}{osc}}
\newcommand{\DLbol}{\down{\DL}{bol}}
\newcommand{\DLlambda}{\DL_{\lambda}}
\newcommand{\DL}{(\delta L/L)}
\newcommand{\DW}{(\delta W/W)}
\newcommand{\DT}{\overline{\Delta T}}
\newcommand{\dnuz}{\delta \nu_0}
\newcommand{\Dnuz}{\Delta \nu_0}
\newcommand{\Dnu}[1]{\Delta \nu_{#1}}
\newcommand{\dnu}[1]{\delta \nu_{#1}}
\newcommand{\DvSun}{\Dv\commaSun}
\newcommand{\Dv}{\down{v}{osc}}
\newcommand{\Fcon}{\down{F}{con}}
\newcommand{\HP}{\down{H}{P}}
\newcommand{\Hrho}{H_{\rho}}
\newcommand{\cP}{\down{c}{P}}
\newcommand{\cms}{\mbox{cm\,s$^{-1}$}}
\newcommand{\commaSun}{\mbox{$_{,\odot}$}}
\newcommand{\cs}{\down{c}{s}}
\newcommand{\kms}{\mbox{km\,s$^{-1}$}}
\newcommand{\lambdabol}{\down{\lambda}{bol}}
\newcommand{\mean}[1]{\langle#1\rangle}
\newcommand{\ms}{\mbox{m\,s$^{-1}$}}
\newcommand{\muHz}{\mbox{$\mu$Hz}}
\newcommand{\nmax}{\down{n}{max}}
\newcommand{\nuac}{\down{\nu}{ac}}
\newcommand{\numax}{\down{\nu}{max}}
\newcommand{\nurot}{\down{\nu}{rot}}
\newcommand{\sigmaRMS}{\down{\sigma}{rms}}
\newcommand{\sigmaAMP}{\down{\sigma}{amp}}
\newcommand{\sigmaPS}{\down{\sigma}{PS}}
\newcommand{\tsub}{\down{t}{sub}}
\newcommand{\vcon}{\down{v}{con}}

\newcommand{\epsEri}{\mbox{$\varepsilon$~Eri}}
\newcommand{\aCenA}{\mbox{$\alpha$~Cen~A}}
\newcommand{\bHyi}{\mbox{$\beta$~Hyi}}

\newcommand{\PSPS}{PS$\otimes$PS}
\newcommand{\Teff}{\down{T}{eff}}

\newcommand{\KB}{KB}
\newcommand{\KBVF}{KBVF}

\newcommand{\Halpha}{\mbox{H$\alpha$}}
\newcommand{\Hbeta}{\mbox{H$\beta$}}
\newcommand{\Hgamma}{\mbox{H$\gamma$}}
\newcommand{\Hdelta}{\mbox{H$\delta$}}
\newcommand{\Hepsilon}{\mbox{H$\epsilon$}}
\newcommand{\CaII}{\mbox{Ca~{\sc ii}}}
\newcommand{\CaK}{\mbox{Ca~{\sc ii}~K}}
\newcommand{\CaHeps}{\mbox{Ca~{\sc ii}~H + \Hepsilon}}
\newcommand{\FeI}{\mbox{Fe~{\sc i}}}
\newcommand{\HI}{\mbox{H~{\sc i}}}
\newcommand{\MgI}{\mbox{Mg~{\sc i}}}
\newcommand{\OI}{\mbox{O~{\sc i}}}

\ifnfsstwo
  \newcommand{\mitbf}[1] {\hbox{\mathversion{bold}$#1$}}
  \newcommand{\rmn}[1] {{\mathrm #1}}
  \newcommand{\itl}[1] {{\mathit #1}}
  \newcommand{\bld}[1] {{\mathbf #1}}
\fi

\ifnfssone
  \newmathalphabet{\mathit}
    \addtoversion{normal}{\mathit}{cmr}{m}{it}
    \addtoversion{bold}{\mathit}{cmr}{bx}{it}
    \newcommand{\mitbf}[1] {\hbox{\mathversion{bold}$#1$}}
    \newcommand{\rmn}[1] {{\mathrm #1}}
    \newcommand{\itl}[1] {{\mathit #1}}
    \newcommand{\bld}[1] {{\mathbf #1}}
\fi

\ifoldfss    
  \newcommand{\mitbf}[1] {\mbox{\boldmath $#1$}}
  \newcommand{\rmn}[1] {{\rm #1}}
  \newcommand{\itl}[1] {{\it #1}}
  \newcommand{\bld}[1] {{\bf #1}}
\fi

\loadboldmathitalic
\loadboldgreek

\title[Stellar oscillations in equivalent width]{Measuring stellar
oscillations using equivalent widths of absorption lines}
\author[T.~R.~Bedding et al.]
       {T.~R.~Bedding,$^1$
	H.~Kjeldsen,$^{2,3}$
	J.~Reetz$^4$
	and B.~Barbuy$^5$\\
	$^1$Chatterton Astronomy Department, School of Physics, University
		of Sydney 2006, Australia {\tt
		(bedding@physics.usyd.edu.au)}\\ 
	$^2$Teoretisk Astrofysik Center, Danmarks Grundforskningsfond\\
	$^3$Institute of Physics and Astronomy, Aarhus University, DK-8000
		Aarhus C, Denmark\\
	$^4$Institut f\"ur Astronomie und Astrophysik der Universit\"at
M\"unchen, Scheinerstr.~1, D-81679 M\"unchen, Germany\\
	$^5$Universidade de S\~ao Paulo, IAG, Depto.\ de Astronomia,
        CP 9638, S\~ao Paulo, 01065-970, Brazil
}
\pubyear{1996}

\begin{document}

\maketitle

\begin{abstract}
Kjeldsen et al.\ (1995\comment{, AJ {\bf 109,} 1313}) have developed a new
technique for measuring stellar oscillations and claimed a detection in the
G subgiant \eBoo.  The technique involves monitoring temperature
fluctuations in a star via their effect on the equivalent width of Balmer
lines.  In this paper we use synthetic stellar spectra to investigate the
temperature dependence of the Balmer lines, \CaII, \FeI, the Mg b feature
and the G~band.  We present a list of target stars likely to show
solar-like oscillations and estimate their expected amplitudes.  We also
show that centre-to-limb variations in Balmer-line profiles allow one to
detect oscillation modes with $\ell\le4$, which accounts for the detection
by Kjeldsen et al.\ of modes with degree $\ell=3$ in integrated sunlight.
\end{abstract}


\section{Introduction}

Measuring stellar oscillations is extremely difficult.  The five-minute
oscillations in the Sun, while rich in their information content, have tiny
amplitudes.  Observing similar oscillations in other stars therefore poses
a great challenge (see {}\citebare{B+G94} for a recent review).  Most
attempts have sought to detect periodic Doppler shifts of spectral lines.
A second method involves measuring the total stellar luminosity, which
varies due to changes in temperature induced by acoustic waves in the
stellar atmosphere.  {}\citeone[hereafter referred to as {}\KBVF]{KBV95-def}
have proposed a new method in which they measure temperature changes via
their effect on the equivalent widths of the Balmer hydrogen lines.  They
presented strong evidence for oscillations in the G~subgiant {}\eBoo\ (see
also {}\citebare{B+K95}), with frequency splittings that were later found
to agree with theoretical models (\citebare{ChDBH95};
Christensen-Dalsgaard, Bedding \& Kjeldsen {}\citeyear{ChDBK95}).


Here we examine the equivalent-width method in more detail.  In
{}\secref{sec.Teff} we use synthetic spectra to calculate the temperature
sensitivity of the Balmer lines and other strong absorption features.  In
{}\secref{sec.targets} we use these results to estimate oscillation
amplitudes for a list of target stars.  In {}\secref{sec.spatial} we
discuss the centre-to-limb behaviour of absorption lines and show that
observing Balmer-line equivalent widths allows one to detect non-radial
oscillation modes with $\ell\le4$.


\section{Temperature dependence of equivalent widths}
	\label{sec.Teff}

The amplitude of an oscillation when measured in equivalent width~($W$) is
related to the temperature fluctuations via
\beq
   \frac{\delta W}{W}
         =  \frac{\partial\ln W}{\partial\ln T} \;
                       \frac{\delta T}{T}.
	  \label{eq.dW}
\eeq
Lines that are highly sensitive to temperature can be used to measure
oscillations, while temperature-insensitive lines are useful as references.
Also, detection sensitivity can be improved by combining observations of
different lines, but only if the temperature dependence is known for each
one.

To determine the function $W(T)$, we shall assume that an oscillating star
can be represented at all times by a model atmosphere in radiative
equilibrium.  A more sophisticated approach would require that the acoustic
waves be treated explicitly (see {}\citebare{Bre75} and
{}\citebare{Fra84}).  With this assumption, we may calculate $W(T)$ by
considering a sequence of normal stellar atmospheres spanning a range of
effective temperatures.  This is best done using theoretical models, since
databases of observed stellar spectra are inhomogeneous with respect to
metallicity and surface gravity, and have uncertainties in $\Teff$ for each
star.  Note that in this paper we are not trying to estimate equivalent
widths exactly.  We are interested in relative changes as a function of
effective temperature (and, in \secref{sec.spatial}, of centre-to-limb
position).

\begin{figure}

\centerline{
\psfig{%
bbllx=147pt,bblly=40pt,bburx=569pt,bbury=453pt,%
figure=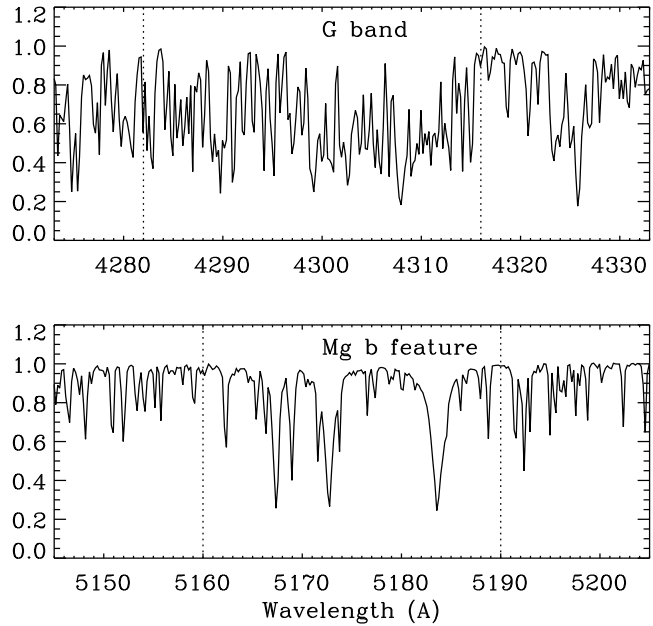,%
width=\the\hsize,scale=100}}

\caption[]{\label{fig.spectra}
Synthetic spectra of the G band and the Mg~b feature for a model atmosphere
having $\Teff=5800$\,K.  Equivalent widths were estimated by measuring the
total flux in the regions enclosed by dashed lines.  }

\end{figure}

We have used the theoretical model GRS88 \cite{FAG93} to calculate spectra
of {}\Halpha, {}\Hbeta, \CaK, the \CaHeps\ blend and the 4383\,\AA\ {}\FeI\
line.  Line profiles for the Balmer transitions were precalculated by
K.~Butler and T.~Sch\"oning (private communication) using VCS theory
{}\cite{VCS70}.  We have also calculated spectra of the G~band and the Mg~b
feature (see {}\figref{fig.spectra}) using the LTE code RAI11 by M.~Spite,
as extended for molecular lines by {}\citeone{Bar82} --- see
{}\citetwo{Bar81}{Bar82} and {}\citeone{CPB91} for descriptions of the
atomic and molecular data bases.  The model atmospheres used in this case
are from Gustafsson et al.\ (unpublished).

For all models we have adopted solar surface gravity and metallicity;
calculations of the~$\Halpha$ line by {}\citeone{FAG93} indicate that
changing these parameters has little effect on equivalent widths.  The
microturbulent velocity was assumed to be 1\,km/s (note this value is not
critical for the strong lines we are considering) and the damping constants
were obtained by fitting to the solar spectra (and also the Arcturus
spectrum, for the G~band and Mg~b).  The models do not include the
chromospheric temperature inversion, but chromospheric effects should have
negliglible influence on the equivalent widths of the metal lines
considered here because only the innermost line cores are formed at
$\log\tau_{5000} < -3.5$.  For the Balmer line, equivalent widths near the
limb {\em are\/} significantly influenced by non-LTE effects in the line
core, as discussed in \secref{sec.spatial.calc} below.

\begin{figure}

\centerline{
\psfig{%
bbllx=27pt,bblly=26pt,bburx=570pt,bbury=751pt,%
figure=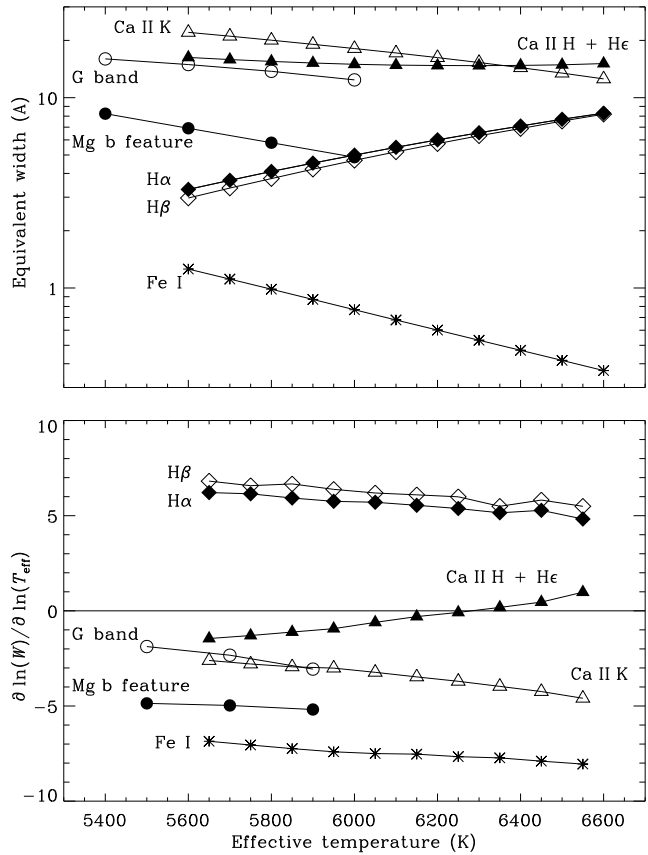,%
width=\the\hsize,scale=100}}

\caption[]{\label{fig.slope}
Equivalent width and slope ($\partial\ln W/\partial\ln \Teff$) for
various lines as a function of stellar effective temperature, based on
model atmosphere calculations.  }

\end{figure}

The upper panel in \figref{fig.slope} shows $W(\Teff)$ as measured from our
synthetic spectra.  The lower panel shows the slope ($\partial\ln
W/\partial\ln\Teff$), which we obtained simply by measuring the differences
between adjacent points in the upper panel.  The slope for the Balmer lines
is 5--7, consistent with the value of 6 adopted by {}\KBVF, which was based
on published models {}\cite{Kur79,Gray92}.  We see that the Mg~b feature is
especially promising for oscillation measurements.  The {}\FeI\ lines are
also useful since, although they are individually weaker, they are numerous
and very sensitive to temperature.


\section{Oscillations in target stars}
	\label{sec.targets}

In the Sun, the amplitudes of individual oscillation modes can vary
considerably over timescales of several days.  However, different modes
vary independently and the maximum amplitude, taken over all modes, stays
roughly constant.  When the solar oscillations are measured in bolometric
luminosity, this amplitude is $\DLbol = 4.1$\,ppm (parts-per-million) ---
see {}\citeone[hereafter referred to as {}\KB]{K+B95-def} and references
therein.

From \KB, equations 5 and 7, amplitudes of solar-like oscillations in other
stars should scale roughly as
\beq
  \DLbol =\frac{L/\Lsol}{(M/\Msol)(\Teff/5777\,{\rm K})} \; 4.1\,{\rm ppm}.
	\label{eq.dL}
\eeq
This luminosity variation is due almost entirely to changes in temperature
(the change in radius is negligible).  Therefore, using $L\propto R^2 T^4$
and {}\eqref{eq.dW} we can write
\beq
   \frac{\delta W}{W}
         =  \frac{\partial\ln W}{\partial\ln \Teff} \;
		\frac{L/\Lsol}{(M/\Msol)(\Teff/5777\,{\rm K})} \;
		1.0\,{\rm ppm}.
	\label{eq.signal}
\eeq

\begin{table*}

\caption[]{\label{tab.targets} Expected oscillations signal 
from observations of H$\alpha$ equivalent width for a sample of bright stars}
\small
\hspace*{-3em}
\begin{tabular}{
r l l r r r r r r r r r r r}
\noalign{\smallskip}
\multicolumn{1}{c}{HR} & 
\multicolumn{1}{l}{Name} & 
\multicolumn{1}{l}{Spectral} & 
\multicolumn{1}{c}{$V\!\!\!$} & 
\multicolumn{1}{c}{$\Teff$} & 
\multicolumn{1}{c}{$L$} & 
\multicolumn{1}{c}{$M$} & 
\multicolumn{1}{c}{$W$} & 
\multicolumn{1}{c}{$\delta W/W\!\!\!$} & 
\multicolumn{1}{c}{$S/N$} & 
\multicolumn{1}{c}{$\numax$   } & 
\multicolumn{1}{c}{$\Dnuz$} & 
\multicolumn{1}{c}{$\nurot\sin i\!\!$} \\
\multicolumn{1}{c}{} & 
\multicolumn{1}{l}{} & 
\multicolumn{1}{l}{Type} & 
\multicolumn{1}{c}{} & 
\multicolumn{1}{c}{(K)} & 
\multicolumn{1}{c}{$(\Lsol)$} & 
\multicolumn{1}{c}{$(\Msol)$} & 
\multicolumn{1}{c}{(\AA)} & 
\multicolumn{1}{c}{(ppm)$\!\!\!$} & 
\multicolumn{1}{c}{} & 
\multicolumn{1}{c}{($\muHz$)} & 
\multicolumn{1}{c}{($\muHz$)} & 
\multicolumn{1}{c}{($\muHz$)} \\
\noalign{\smallskip}
\hline
\noalign{\smallskip}
2943 & $\alpha$ CMi        &  F5IV-V          & $ 0.38$ &  6500 &     7.1 &  1.60 &    7.7 &     20 &  10.7 &   1050 &     56 & $   0.7$ \\
5235 & $\eta$ Boo          &  G0IV            & $ 2.68$ &  6050 &     9.5 &  1.60 &    5.2 &     32 &   5.0 &    600 &     36 & $   1.1$ \\
8974 & $\gamma$ Cep        &  K1III-IV        & $ 3.21$ &  4850 &    13.6 &  1.45 &    1.2 &     83 &   4.8 &    200 &     14 & $<   0.7$ \\
 544 & $\alpha$ Tri        &  F6IV            & $ 3.41$ &  6350 &    10.8 &  1.65 &    6.9 &     31 &   3.9 &    650 &     39 & $   7.8$ \\
5459 & $\alpha$ Cen A      &  G2V             & $-0.01$ &  5770 &     1.5 &  1.09 &    4.0 &      8 &   3.7 &   2300 &    106 & --~ \\
\noalign{\medskip}
6380 & $\eta$ Sco          &  F3III-IVp       & $ 3.33$ &  6650 &     9.6 &  1.60 &    8.7 &     25 &   3.7 &    850 &     48 & $  14.7$ \\
3775 & $\theta$ UMa        &  F6IV            & $ 3.17$ &  6450 &     9.4 &  1.70 &    7.5 &     25 &   3.7 &    800 &     46 & $   1.1$ \\
8961 & $\lambda$ And       &  G8III-IV        & $ 3.82$ &  4900 &    14.0 &  1.50 &    1.3 &     81 &   3.7 &    200 &     14 & $<   0.8$ \\
6212 & $\zeta$ Her         &  G0IV+G7V        & $ 2.90$ &  5800 &     5.8 &  1.30 &    4.2 &     27 &   3.3 &    700 &     42 & $<   1.0$ \\
 --~ & Sun 1  pc           &  G2V             & $-0.15$ &  5777 &     1.0 &  1.00 &    4.0 &      6 &   3.0 &   3050 &    135 & $   0.5$ \\
\noalign{\medskip}
 566 & $\chi$ Eri          &  G8III           & $ 3.70$ &  5250 &     9.9 &  1.65 &    2.2 &     45 &   2.8 &    350 &     24 & --~ \\
7957 & $\eta$ Cep          &  K0IV            & $ 3.43$ &  5100 &     7.9 &  1.50 &    1.7 &     42 &   2.6 &    350 &     24 & $<   1.1$ \\
5986 & $\theta$ Dra        &  F8IV            & $ 4.01$ &  6250 &     7.8 &  1.50 &    6.3 &     26 &   2.4 &    750 &     45 & $   2.6$ \\
  98 & $\beta$ Hyi         &  G2IV            & $ 2.80$ &  5800 &     2.7 &  0.99 &    4.1 &     17 &   2.1 &   1150 &     64 & --~ \\
8665 & $\xi$ Peg           &  F6III-IV        & $ 4.19$ &  6300 &     6.3 &  1.40 &    6.7 &     22 &   1.9 &    900 &     52 & $   0.8$ \\
\noalign{\medskip}
7602 & $\beta$ Aql         &  G8IV            & $ 3.71$ &  5250 &     5.6 &  1.40 &    2.1 &     30 &   1.8 &    550 &     33 & $<   1.3$ \\
7061 & 110 Her             &  F6V             & $ 4.19$ &  6450 &     6.3 &  1.55 &    7.5 &     19 &   1.7 &   1100 &     60 & $   1.6$ \\
 963 & $\alpha$ For        &  F8V+G7V         & $ 3.90$ &  6250 &     4.0 &  1.30 &    6.3 &     16 &   1.5 &   1300 &     69 & --~ \\
 458 & $\upsilon$ And      &  F8V             & $ 4.09$ &  6200 &     4.4 &  1.30 &    5.9 &     17 &   1.5 &   1150 &     62 & $   1.0$ \\
3579 & ~---                &  F5V+G5V         & $ 4.11$ &  6550 &     3.8 &  1.13 &    8.0 &     15 &   1.5 &   1400 &     77 & $   3.9$ \\
\noalign{\medskip}
8430 & $\iota$ Peg         &  F5V             & $ 3.76$ &  6550 &     3.7 &  1.30 &    8.0 &     13 &   1.5 &   1650 &     84 & $   1.1$ \\
1136 & $\delta$ Eri        &  K0IV            & $ 3.54$ &  5100 &     3.2 &  1.10 &    1.7 &     24 &   1.4 &    650 &     40 & $<   1.7$ \\
1543 & $\pi^3$ Ori         &  F6V             & $ 3.09$ &  6650 &     2.5 &  1.30 &    8.5 &      8 &   1.3 &   2600 &    118 & $   3.3$ \\
5404 & $\theta$ Boo        &  F7V             & $ 4.05$ &  6300 &     3.8 &  1.30 &    6.7 &     14 &   1.3 &   1450 &     75 & $   4.8$ \\
4540 & $\beta$ Vir         &  F9V             & $ 3.61$ &  6150 &     2.8 &  1.20 &    5.7 &     12 &   1.3 &   1650 &     83 & $   0.5$ \\
\noalign{\medskip}
6623 & $\mu$ Her           &  G5IV            & $ 3.42$ &  5550 &     2.2 &  1.00 &    3.0 &     15 &   1.3 &   1150 &     64 & $   2.8$ \\
5933 & $\gamma$ Ser        &  F6V             & $ 3.85$ &  6400 &     3.3 &  1.30 &    7.1 &     12 &   1.2 &   1700 &     85 & $   1.2$ \\
8969 & $\iota$ Psc         &  F7V             & $ 4.13$ &  6300 &     3.2 &  1.25 &    6.5 &     13 &   1.1 &   1600 &     81 & $   0.9$ \\
 799 & $\theta$ Per        &  F8V             & $ 4.12$ &  6350 &     2.7 &  1.25 &    6.9 &     10 &   0.9 &   1950 &     95 & $   1.0$ \\
 937 & $\iota$ Per         &  G0V             & $ 4.05$ &  6000 &     2.4 &  1.15 &    5.0 &     12 &   0.9 &   1700 &     85 & $<   1.6$ \\
\noalign{\medskip}
1983 & $\gamma$ Lep        &  F6V             & $ 3.60$ &  6450 &     1.8 &  1.25 &    7.3 &      7 &   0.8 &   3100 &    135 & $   2.3$ \\
5460 & $\alpha$ Cen B      &  K1V             & $ 1.33$ &  5350 &     0.5 &  0.90 &    2.4 &      4 &   0.7 &   4450 &    179 & --~ \\
6927 & $\chi$ Dra          &  F7V             & $ 3.57$ &  6350 &     1.8 &  1.50 &    6.9 &      6 &   0.7 &   3450 &    139 & $   2.2$ \\
7665 & $\delta$ Pav        &  G6-8IV          & $ 3.56$ &  5500 &     1.1 &  0.90 &    2.9 &      9 &   0.7 &   2050 &    100 & --~ \\
7936 & $\psi$ Cap          &  F4V             & $ 4.14$ &  6600 &     1.8 &  1.25 &    8.2 &      6 &   0.6 &   3300 &    142 & $   8.1$ \\
\noalign{\medskip}
 219 & $\eta$ Cas          &  G0V+dM0         & $ 3.45$ &  6100 &     1.1 &  1.15 &    5.4 &      5 &   0.6 &   3750 &    154 & $<   1.4$ \\
 --~ & Sun 10 pc           &  G2V             & $ 4.85$ &  5777 &     1.0 &  1.00 &    4.0 &      6 &   0.3 &   3050 &    135 & $   0.5$ \\
 509 & $\tau$ Cet          &  G8V             & $ 3.50$ &  5600 &     0.4 &  0.90 &    3.3 &      3 &   0.3 &   5650 &    218 & $   0.7$ \\
1084 & $\varepsilon$ Eri   &  K2V             & $ 3.73$ &  5180 &     0.3 &  0.85 &    2.0 &      3 &   0.2 &   5350 &    206 & $<   5.4$ \\
\noalign{\smallskip}
\hline 
\end{tabular}
\end{table*}

 \Eqref{eq.signal} gives an estimate for the oscillation signal expected
from equivalent-width measurements of a spectral line.  The photon noise,
on the other hand, scales as the square root of the number of photons
detected in the line.  Hence, assuming photon noise to be the dominant
noise source, the signal-to-noise ratio for a fixed observing time will
scale as:
\beqa
 S/N     & \propto & \frac{\partial\ln W}{\partial\ln \Teff} \; 
      \frac{L \sqrt{W} \, 10^{-0.2 m}}{M\Teff},
	\label{eq.SN}
\eeqa
where~$m$ is the stellar magnitude at the relevant wavelength.

We have applied {}\eqsref{eq.signal}{eq.SN} to a sample of bright stars
that are likely to undergo solar-like oscillations.  {}\Tabref{tab.targets}
shows fundamental parameters and expected oscillation characteristics for
the stars, which were selected from {\em The Bright Star Catalogue\/}
{}\cite{Hoffleit} to have $V < 4.2$ and $0.35 < B-V < 1.2$.  A further
restriction to include only main-sequence stars and subgiants was made on
the basis of density, as described below.  We also omitted known variables
(e.g., $\delta$~Scuti stars) and close binaries having nearly-equal
components.  For the remaining binaries we give the parameters for the
primary only (except $\aCen$, whose components are listed separately).

For the Sun, $\aCen$~A and~B, Procyon ($\alpha$~CMi), $\eBoo$, $\bHyi$ and
$\epsEri$ we adopted the same fundamental parameters as used in \KB\ and
\KBVF.  For the other stars, we estimated effective temperatures using
$\Teff \simeq 11000\,{\rm K} / (B-V + 1.24)$, used parallaxes from {\em The
Bright Star Catalogue\/} and took bolometric corrections from
{}\citeone{LangData}.  We made no reddening corrections except for HR~1543,
which has $E(B-V)=0.03$.  Masses were estimated using stellar models in
Fig.~1 of {}\citeone{BFB93}, which assumes solar metallicity, OPAL
opacities and convective overshoot.

For each star we used {}\eqref{eq.signal} to calculate the oscillation
signal expected from observations of equivalent-width fluctuations
of~$\Halpha$.  Computing similar results for other lines is straightforward
using the data in {}\figref{fig.slope}.  The predictions for $\eBoo$ and
the Sun agree roughly with the Balmer-line observations of {}\KBVF,
although the amplitude scale in {}\tabref{tab.targets} may need revision
once more observations become available (we are currently analysing
high-quality data from $\aCenA$, Procyon and the Sun).  However, the
relative amplitudes of the stars should not change drastically unless
{}\eqref{eq.dL}, which is based on theoretical models, is found to be
incorrect.

The stars in {}\tabref{tab.targets} are ordered according to $S/N$.  This
is the signal-to-noise ratio for a fixed observing time, arbitrarily
normalized to give $S/N=3$ for the Sun at a distance of 1\,pc.  How much
observing time would this require?  Based on the KBVS observations of
{}\eBoo, a total of $1.4\times10^{10}$ photons/\AA\ are required to produce
a photon-noise level of 10\,ppm.  This allows us to estimate the observing
time required to achieve a given $S/N$.  For example, a 2.5-metre telescope
collecting photons with a total efficiency of 5\% would take about 43 hours
to achieve the $S/N$ given for each star in {}\tabref{tab.targets}.  This
estimate is based on observations at $\Halpha$ only -- simultaneous
measurement of other spectral lines would improve matters.  Also, we have
neglected $1/f$ noise sources, such as instrumental drift, which will be
important for stars oscillating at low frequencies (i.e., sub-giants).

The ranking of targets in {}\tabref{tab.targets} may differ for the other
observing techniques (luminosity and velocity) and can be calculated using
the formulae in \KB.  The Table also includes rough estimates for the
frequency of maximum oscillation amplitude~$\numax$ and the large frequency
separation~$\Dnuz$, computed by scaling from the Sun (see {}\KB).  These
frequencies will be important when choosing targets and designing observing
programs.  The large separation depends on the average density of the star
and the table only includes stars for which $\Dnuz > 1/({\rm day}) =
11.57\,\muHz$, in order to restrict the sample to main-sequence stars and
subgiants.

The table also shows the quantity
\beqa
 \nurot\sin i & = & v\,\sin i / (2\pi R) \\
              & \simeq & 0.23\,\muHz\, \frac{v\,\sin i/\kms}{R/\Rsol},
\eeqa
where~$\nurot$ is rotation frequency of the star.  Components of an
oscillation mode that is split by stellar rotation will be separated by
approximately~$\nurot$ (see below).

Finally, it is worth mentioning the double-lined spectroscopic binary
$\alpha$~Equ (HR~8131), whose components (G5~III and A5~V) are slightly too
faint to be included in {}\tabref{tab.targets}.  The masses and
luminosities of both components have been derived from interferometric and
spectroscopic observations by {}\citeone{AMV92}, making this system
particularly valuable for testing stellar models.


\section{Mode sensitivity}	\label{sec.spatial}

\subsection{Calculations}	\label{sec.spatial.calc}

\begin{figure}

\centerline{
\psfig{%
bbllx=102pt,bblly=26pt,bburx=569pt,bbury=468pt,%
figure=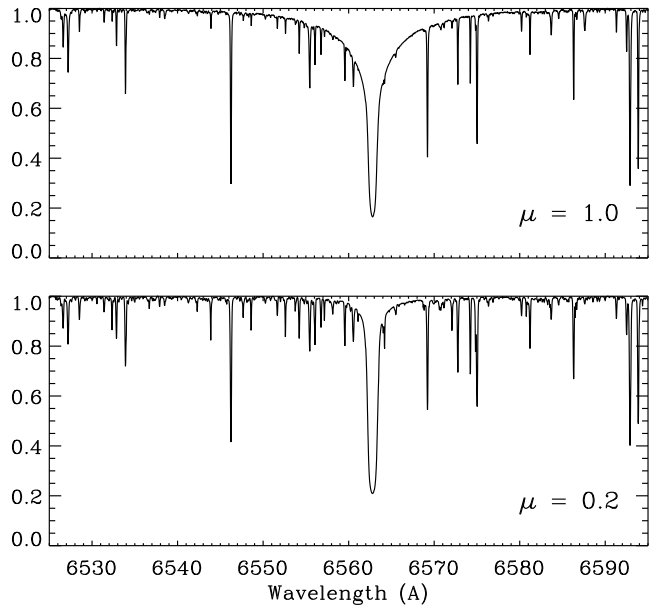,%
width=\the\hsize,scale=100}}

\caption[]{\label{fig.kpno}
Observed $\Halpha$ spectra of the Sun at the centre and edge of the disk,
taken from {}\citeone{B+T72}.  The wings of the $\Halpha$ line are much
weaker in the lower diagram, whereas the metal lines and the core of
$\Halpha$ show little variation. }

\end{figure}

From equivalent-width observations of the daytime sky, \KBVF\ detected
solar oscillations with degrees $\ell=0$, 1, 2 and~3.  The detection of
$\ell=3$ was puzzling, given that the the observations did not resolve the
solar disk.  The answer is straightforward: the Balmer lines are much
weaker at the edge of the solar disk than at the centre (see
{}\figref{fig.kpno}), so their integrated profiles are strongly weighted
towards the disk centre.  This has lead us to investigate in detail the
sensitivity of the equivalent-width method to modes with different values
of~$\ell$.

The calculation involves integrating the oscillations over the stellar
surface, including the effects of limb darkening
{}\cite{ChD+G82,HWS94,RMZ95}.  The oscillations of a star can be described
in terms of spherical harmonics $Y_\ell^m(\theta,\phi)$.  If the star is
rotating slowly, there is no preferred direction and we are free to choose
a coordinate system with the polar axis pointed towards the observer.  In
this case, all integrations of spherical harmonics will be zero unless $m =
0$.  Even for stars in which rotation is important, the following analysis
is still useful, as discussed at the end of this section.

For $m = 0$, the spatial response function for each~$\ell$ (the ratio of
the observed to the actual amplitude) is
\beq
 S_{\ell} = 2 \sqrt{2 \ell + 1} \int^1_0 P_\ell(\mu) \, I(\mu) \, F(\mu) \,
       \mu \, d\mu. \label{eq.Sl}
\eeq
Here $\mu =\cos\theta$, $P_\ell(\mu)$ is the Legendre polynomial of
degree~$\ell$, $I(\mu)$ is the centre-to-limb variation of continuum
intensity (classical limb darkening) and $F(\mu)$ is the centre-to-limb
variation of the oscillation signal.

The function $I(\mu)$ for the Sun is well approximated by
\beq
   I(\mu) = 1 - u_2 (1 - \mu) - v_2 (1 - \mu^2),
\eeq
where the coefficients vary with wavelength \cite{Allen}.  The function
$F(\mu)$ depends on how the observations are made.  If the oscillations are
observed via the Doppler shift of a spectral line then $F(\mu) = \mu$,
corresponding to the projection of a nearly-radial pulsation along the line
of sight.  For observations in intensity there is no projection effect (to
a good approximation) and $F(\mu) = 1$.  If the oscillations are observed
in equivalent width then we set $F(\mu) = W(\mu)$, the spatial variation of
equivalent width from centre to limb.

In principle, we could determine the function $W(\mu)$ from
spatially-resolved spectra of the Sun.  However, at visible wavelengths it
seems that only two have been published: the {\em Preliminary Kitt Peak
Photoelectric Atlas\/} {}\cite{B+T72}, which has spectra at $\mu = 1.0$ and
$\mu = 0.2$ (see {}\figref{fig.kpno}), and the atlas by {}\citeone{DRN73},
which contains a single spectrum at $\mu = 1.0$.

\begin{figure}

\centerline{
\psfig{%
bbllx=90pt,bblly=154pt,bburx=522pt,bbury=538pt,%
figure=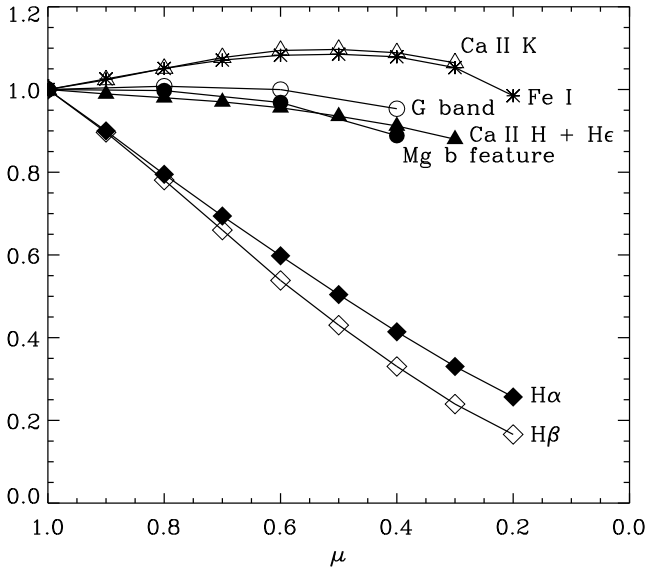,%
width=\the\hsize,scale=100}}

\caption[]{\label{fig.limb}
Centre-to-limb variation of equivalent width from solar-temperature model
atmospheres, normalized to unity at the centre of the disk.  }

\end{figure}

To estimate $W(\mu)$ at all values of $\mu$, we have therefore used the
models described in {}\secref{sec.Teff} to produce intensity spectra at
different points on a stellar disk.  The results for a star with solar
temperature, normalized to the disk centre, are shown in
{}\figref{fig.limb}.  For the Balmer line cores, non-LTE effects are
important and the models significantly underestimated the area of the line
cores near the limb.  To account for this, we assumed the difference
between the actual equivalent width and that predicted by the GRS88 model
increases linearly with decreasing $\mu$.  Based on the measurements of
$W(\mu)$ at two points on the solar disk ($\mu=1.0$ and 0.2), we could
correct the GRS88 equivalent widths at all intermediate points (the
correction at $\mu=0.3$ was 35\% for \Halpha\ and 18\% for \Hbeta).  The
final values are shown in \figref{fig.limb}.

For those lines that were computed using the GRS88 model, we have repeated
the calculations using the semi-empirical model of {}\citeone{H+M74}.  For
the Balmer lines, we again corrected the equivalent widths for the missing
core area.  The resulting $W(\mu)$ profiles for all lines are similar to
those in \figref{fig.limb}, to within 10\%.

One other comparison with observation is possible.  Quoting the unpublished
{\em Michigan Atlas\/}, {}\citeone{Hol67} gives equivalent widths at
$\mu=1.0$ and $\mu=0.3$ for many lines, including 5172\,\AA\ $\MgI$ and
4383\,\AA\ $\FeI$ (but note that his values for $\CaII$~H~\&~K and $\HI$
come from model calculations).  For these lines, the ratio
$W(\mu=0.3)/W(\mu=1.0)$ given by {}\citename{Hol67} agrees with our
calculations to within 10\%.  It would clearly be desirable to have
detailed centre-to-limb measurements of all these strong lines, similar to
those of the 7774\,\AA\ $\OI$ triplet by {}\citeone{Alt68} and
{}\citeone{King+B95}.

Our general conclusion is that the Balmer lines are strongly limb darkened,
while the other lines have almost constant strength from centre to limb.
For our purposes, it is sufficient to approximate $F(\mu)$ for each
spectral line using a linear function:
\begin{equation}
   F(\mu ) = 1 - c (1 - \mu).
	\label{eq.F}
\end{equation}
By fitting a straight line to each set of points in {}\figref{fig.limb}
(giving more weight to the central portion of the disk, since this
contributes the greatest area), we obtained the following values for~$c$:
0.95 for $\Halpha$, 1.0 for $\Hbeta$ (which we also adopt for $\Hgamma$ and
$\Hdelta$), $-0.2$ for {}\CaK, 0.1 for the {}\CaHeps\ blend, $-0.2$ for
\FeI, 0.05 for the G~band and 0.15 for the Mg~b feature.  

Our Balmer-line models indicate that $c$ varies quite slowly with effective
temperature ($c\propto\Teff^{-0.5}$), so the results given here apply to a
broad class of stars.

Having obtained approximations for $I(\mu)$ and $F(\mu)$, we can now
integrate {}\eqref{eq.Sl}.  Results for the five lowest-degree modes
($\ell=0,\dots,4$) are:
\beqa
\left( \begin{array}{c}
S_0\\
S_1\\
S_2\\
S_3\\
S_4
\end{array} \right)
&=&
\left( \begin{array}{cccc}
       1        & \frac{2}{3}      & \frac{1}{2}      & \frac{2}{5}
								\\[1.0ex]
\frac{2}{\surd3}& \frac{\surd3}{2} & \frac{2\surd3}{5}& \frac{1}{\surd3}
								\\[1.0ex]
\frac{\surd5}{4}& \frac{4}{3\surd5}& \frac{\surd5}{4} & \frac{8}{7\surd5}
								\\[1.0ex]
       0        & \frac{\surd7}{12}& \frac{4}{5\surd7}& \frac{\surd7}{8}
								\\[1.0ex]
-\frac{1}{8}    &       0          & \frac{3}{32}     & \frac{16}{105}	 
\end{array} \right)
\times
\nn\\
& & 
\left( \begin{array}{ccc}
1-c & c-1  & c-1 \\
c   & 1-2c &  -c \\
0   & c    & 1-c \\
0   & 0    & c 
\end{array} \right)
\times
\left( \begin{array}{c}
1 \\
u_2 \\
v_2
\end{array} \right)
			\label{eq.matrix}
\eeqa
This matrix product, which should be evaluated from right to left, can be
used to estimate the mode sensitivity of each observing method as a
function of wavelength.  Note that {}\eqref{eq.matrix} can be used for
velocity measurements by setting $c=1$ and for intensity measurements by
setting $c=0$.

\begin{figure}

\centerline{
\psfig{%
bbllx=122pt,bblly=19pt,bburx=468pt,bbury=751pt,%
figure=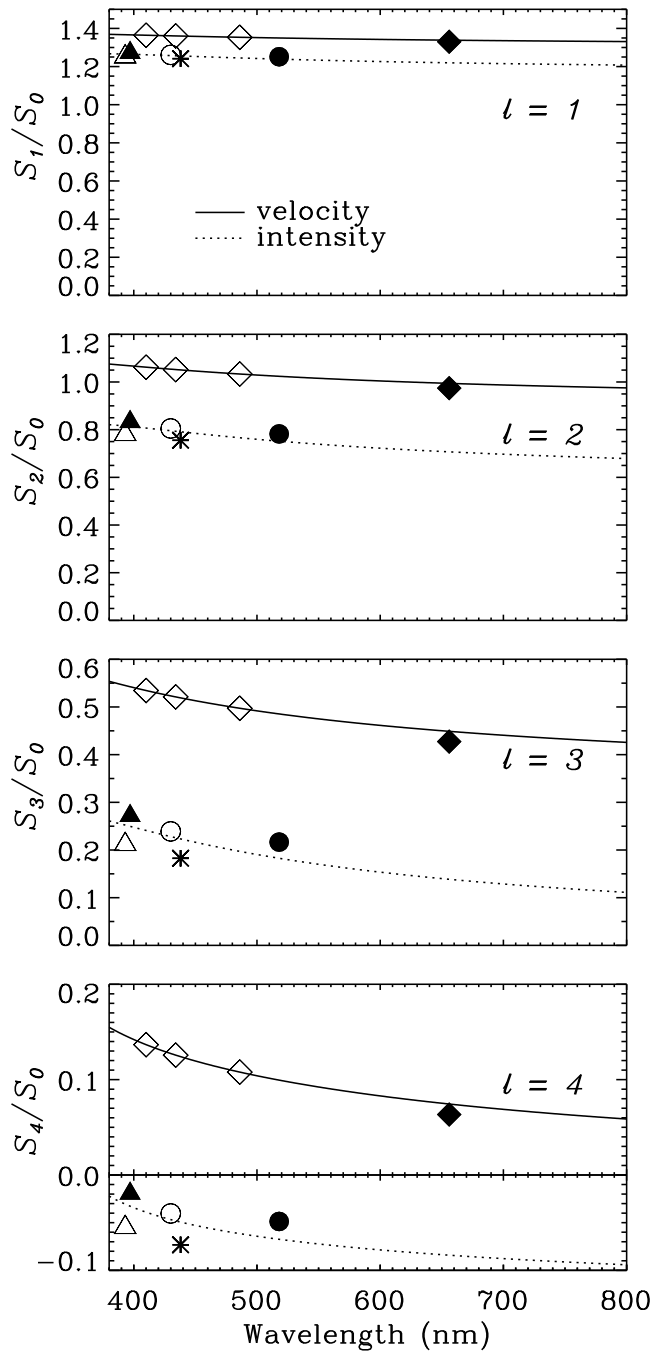,%
width=\the\hsize,scale=100}}

\caption[]{\label{fig.spatial}
Sensitivity of observing methods as a function of wavelength.  The solid
line is for observations in velocity and the dashed line for intensity.
Individual points indicate equivalent-width observations in different
spectral features, with symbols having the same meaning as in
{}\figsref{fig.slope}{fig.limb} (but with the addition of $\Hgamma$ and
$\Hdelta$, shown as open diamonds).  Negative values of $S_{\ell}$ mean the
oscillations will appear to have reversed phases.  }

\end{figure}

\subsection{Results and discussion}

 \Figref{fig.spatial} shows the sensitivities as calculated from
{}\eqref{eq.matrix}, expressed relative to $\ell=0$.  In the case of
velocity and intensity observations (solid and dotted curves), there are
published solar oscillation data with which to compare the calculations:
\ben

 \item Observations in velocity from the South pole using the Na~{\sc i}
line (590\,nm) by {}\citeone{GFP80} give $S_1/S_0 = 1.5$, $S_2/S_0 = 1.2$,
$S_3/S_0 = 0.6$ and $S_4/S_0 = 0.2$ (the scatter is about $\pm0.2$).

 \item Observations in bolometric intensity from the Solar Maximum Mission
by \citeone{W+H83a} give $S_1/S_0 = 1.25$ and $S_2/S_0 = 0.83$ (scatter
about $\pm0.06$).

 \item Observations in luminosity with the IPHIR green channel (500\,nm) by
{}\citeone{T+F92} give $S_1/S_0 = 1.09$ and $S_2/S_0 = 0.79$ (scatter about
$\pm0.07$).

\een
In each case, there is good agreement with our results.

Turning to equivalent-width measurements, we see from \figref{fig.spatial}
that the Balmer lines give a response similar to that of velocity
measurements.  This follows from their strong centre-to-limb variation and
is consistent with the solar observations by {}\KBVF\ --- in particular, we
can now explain their detection of $\ell=3$.  The other absorption lines
considered here have much weaker centre-to-limb variations ($c\simeq 0$;
see {}\figref{fig.limb}) and so have spatial responses similar to that of
intensity measurements.  This suggests the possibility of determining
the~$\ell$ value of a given mode using simultaneous observations in several
absorption lines, which might be interesting for studies of $\delta$~Scuti
variables {}\cite{Mat93}.

Finally, we consider a rotating star.  A mode with a particular $(n,\ell)$
will split into a multiplet having $m=-\ell,\ldots, \ell-1, \ell$, with the
components being separated by about~$\nurot$ (the oscillation power summed
over these peaks is conserved).  If the star is seen pole-on, only modes
with $m=0$ are observable and the analysis of this section holds exactly:
we would not notice that the star is rotating.  At the other extreme, in a
star seen from near the equator (e.g., the Sun) the modes with $|\ell-m|$
odd will have zero amplitude because the Legendre function is antisymmetric
around the equator.  For $\ell=1$ we would therefore only see two modes
($m=\pm 1$).  The total power will be the same as for a non-rotating star,
so the amplitudes of these two modes will each be multiplied by
$1/\sqrt{2}$ relative to the non-rotating case.  For $\ell=2$ we would see
three modes ($m=-2, 0$ and~2) with relative amplitudes of $\sqrt{3/2}$, 1/2
and $\sqrt{3/2}$.  Other cases can readily be calculated using the formulae
given by {}\citeone{ChD94}.

\section*{Acknowledgments}

We are grateful to the following for useful discussions: Richard Altrock,
Lawrence Cram, Bob Fosbury, Jeremy King, Barry LaBonte, Bill Livingston,
Ruth Peterson and Dick White.  We thank the referee for some important
suggestions and Robert Kurucz for supplying the {\em Preliminary Kitt Peak
Photoelectric Atlas\/} in electronic form.

\bibliography{dummy}
\bibliographystyle{astrop-bib-simple}
\bsp

\end{document}